\documentclass[a4pager,12pt]{article}
\usepackage{mathrsfs}
\usepackage{amssymb}
\usepackage{txfonts}
\usepackage{amsfonts}
\title{Effective Field Theory, Renormalizability and Extra Dimensions}
\author{Bei Jia$^{1,2}$ \\{\footnotesize $^1$\textsl{Institute of Modern Physics, Chinese Academy of Sciences,
P.O.Box 31 Lanzhou, 730000, China}}\\{\footnotesize
$^2$\textsl{Graduate University of Chinese Academy of Sciences,
Beijing, 100080, China}}}
\date{}
\begin{document}
\maketitle \abstract We discuss in this paper two ways of defining
the concept of ``effective field theory'': effective field theory
defined by low energy effectiveness and effective field theory
defined by 4D effectiveness out of higher dimensions. We argue that
these two views are actually equivalent, that effective field
theories at low energy can in fact be regarded as field theories of
higher dimensions confined on a 4D spcaetime. We examine this idea
through comparing two different regularization schemes: Momentum
Cutoff and Dimensional Regularization, and through analyzing how
fields can be localized on branes.

\section{Introduction}
Quantum field theories are nowadays usually described as ``effective
field theories'' (EFT) which are only valid below some energy
cutoffs [1$\sim$6]. It is generally believed that we may need a
completely new theory to describe the physics beyond those energy
levels, such as string theory. Also, from an EFT point of view,
renormalizability, which used to be a standard benchmark for
acceptable quantum field theories not very long ago, is no longer
relevant. On the other hand, since the early works of Kaluza and
Klein [7], the concept of extra dimensions has been broadly consumed
by physicists. Now the interests have been shifted from the
traditional Kaluza-Klein type to the so called ``brane world''
picture, in which some fields (like SM fields) are localized at a
brane while other fields (such as gravitation) can propagate in more
dimensions. Numerous models of this kind have been proposed for
different purposes, such as large extra dimensions models like
Arkani-Hamed, Dimopoulos and Dvali (ADD) scenario [8, 9] and warped
extra dimensions models like Randall-Sundrem (RS) models [10, 11].

One might wonder whether the low-energy EFT has any relation with
the 4D EFT. This is the main topic of this paper and we are trying
to answer this question, at least providing a possible clue. We
shall first compare two different regularization schemes in
renormalization: Momentum Cutoff and Dimensional Regularization, in
order to look in to the relationship between effective field theory
with an energy cutoff and effective field theory in four dimensions.
Then in section 3 we move on to discuss the localization of fields
on a 3-brane, in which we will see the non-zero modes of higher
dimensional fields can leave the brane if they reach a high energy
level. Thus the concept of low energy effectiveness and 4D
effectiveness can also be related in this way.

\section{Regularization and Renormalization}
The formal process of renormalization falls into two parts:
regularization and subtraction. Basically we first ``regulate''
divergence in the momentum integral, and then we bring in some
``counter-terms'' to remove the divergence [12]. In renormalization
we are dealing with momentum integrals like

\begin{equation} \label{eps}
I=\int_{0}^{\infty}d^{4}kF(k)
\end{equation}

Usually there are two common ways of regularization: Momentum Cutoff
(MC) and Dimensional Regularization (DR). In the scheme of Momentum
Cutoff, we change the upper limit of the integral in Equation (1)
from  $\infty$ to a momentum cutoff (usually very large) $\Lambda$

\begin{equation} \label{eps}
I\rightarrow I_{\Lambda}=\int_{0}^{\Lambda}d^{4}kF(k)
\end{equation}\\
where $I_{\Lambda}$ is convergent, and becomes $I$ in the limit of
$\Lambda\rightarrow \infty$. Generally this modified integral can be
calculated as

\begin{equation} \label{eps}
I_{\Lambda}=A(\Lambda)+B+C(\frac{1}{\Lambda})
\end{equation}\\
We can see when we perform the limit of $\Lambda\rightarrow\infty$,
we will have $C(1/\Lambda)\rightarrow 0$, $B$ remains the same and
$A(\Lambda)\rightarrow \infty$, which represents the divergence of
the original integral of $I$. This momentum cutoff $\Lambda$ can
actually be described as the energy level under which the field
theory we are using here is effective.

On the other hand, traditionally we start the Dimensional
Regularization scheme with the modification of Equation (1)

\begin{equation} \label{eps}
I\rightarrow I_{D}=\int_{0}^{\infty}d^{D}kF(k)
\end{equation}\\
which means we are shifting the integral from a four dimensional one
to a $D$ dimensional one. It is important to notice that
traditionally we start with a four dimensional field theory, and
only shift to higher dimensions in the process of DR. We will come
back to this later. By defining $\epsilon=4-D$, we can rewrite the
integral as

\begin{equation} \label{eps}
I_{D}=A(\epsilon)+B+C(\frac{1}{\epsilon})
\end{equation}\\
similarly we will have under the limit of $D\rightarrow 4
(\epsilon\rightarrow 0)$, that $A(\epsilon)\rightarrow 0$, $B$
remains the same and $C(1/\epsilon)\rightarrow \infty$. Again we
``parameterized'' the divergence of the original integral, in order
to remove it in the next step of renormalizaition.

Those are the formal ways to perform MC and DR. We would like to
point out that rather than starting with a four dimensional field
theory and then perform DR in the process of renormalization, we
should begin with a $D$ dimensional field theory. By this we mean we
should not wait until we meet the infinity of the momentum integral
and then artificially increase the dimensions. Rather, we should
start our theory in a $D$ dimensional spacetime and then shift it to
4D in order to have an effective theory. The great usefulness of DR
then can be interpreted as following: in order to have a finite
effective theory we have to ``confine'' our starting theory (which
is $D$ dimensional) to 4D spacetime. This point of view suggests
that the divergence of the usual 4D field theory may be regarded as
the effect of higher dimensions.

Let us consider a $D$ dimensional scalar field $\phiup(x_\mu,y_n)$
in a $D$ dimensional Minkowski spcaetime, where $x_\mu\:
(\mu=0,1,2,3)$ denotes the four dimensional coordinates, while
$y_n\: (n=4,5,...,D-1)$ are the coordinates of extra dimensions. We
consider the action

\begin{equation} \label{eps}
S=\int d^{4}x\: d^{D-4}y\: [\frac{1}{2}(\partial _A
\phiup(x_\mu,y_n))^2-\frac{m^2}{2}\phiup^2(x_\mu,y_n)-\frac{\lambdaup}{4!}\phiup^4(x_\mu,y_n)]
\end{equation}\\
where $A$ denotes all the $D$ dimensional coordinates. In order to
get the four dimensional action, we formally integrate over $y_n$

\begin{equation} \label{eps}
S=\int d^{4}x\: [\frac{1}{2}Z_{1}(\partial _\mu
\phiup_{(4)}(x_\mu))^2-\frac{m^2}{2}Z_{2}\phiup_{(4)}^2(x_\mu)-\frac{\lambdaup}{4!}Z_{3}\phiup_{(4)}^4(x_\mu)]
\end{equation}\\
where $Z_{1}$, $Z_{2}$ and $Z_{3}$ are from the integration over
$y_n$, and $\phiup_{(4)}(x_\mu)$ is the result of the integration,
which is a four dimensional scalar field. We can see the four
dimensional effective lagrangian is

\begin{equation} \label{eps}
\mathscr{L}_{eff}=\frac{1}{2}Z_{1}(\partial _\mu
\phiup_{(4)}(x_\mu))^2-\frac{m^2}{2}Z_{2}\phiup_{(4)}^2(x_\mu)-\frac{\lambdaup}{4!}Z_{3}\phiup_{(4)}^4(x_\mu)
\end{equation}\\
This reminds us the traditional argument about the renormalization
of four dimensional $\phiup^4$ theory, in which we have the
``rescaled lagrangian''

\begin{equation} \label{eps}
\mathscr{L}=\frac{1}{2}Z(\partial _\mu
\phiup_{r}(x_\mu))^2-\frac{m^2_0}{2}Z\phiup_{r}^2(x_\mu)-\frac{\lambdaup_0}{4!}Z^{2}\phiup_{r}^4(x_\mu)
\end{equation}\\
where $\phiup^{2}=Z\phiup_{r}^2$, and $m_0$ and $\lambdaup_0$ are
the ``bare parameters''. We can see that, instead of referring the
divergence to the ``bare parameters'', we may say the divergency in
the four dimensional effective lagrangian might come from $Z_{1}$,
$Z_{2}$ and $Z_{3}$, which are the effect of the extra dimensions.

There are many problems about constructing field theories in higher
dimensions [13] and we will not go into this here. We will focus on
the concept of renormalizability. It is important to notice that in
[14] the nonperturbative renormalizability of higher dimensional
gauge field theories is discussed using the functional RG, which
reaches a similar idea with ours. We should mention that the idea of
effective field theory comes from the original work of Wilson about
renormalization group [1]. Traditionally a field theory is called
renormalizabale if the divergent terms in Equations (3) and (5) can
be practically removed by some sort of subtraction. We think that
this renormalizability is only valid in the view of an effective
theory [14, 15] --- when we reach the energy cutoff, our traditional
four dimensional theories lost their effectiveness. Maybe we need
brand new physics like string theory, or we can argue that the loss
of effectiveness is the result of both higher energy level and the
effect of higher dimensions. When we reaches a rather high energy
level, the extra dimensions show up and particles which are used to
be confined in 4D spacetime may escape to those extra dimensions,
and the 4D effective description of field theories is no longer
valid. This leads to the main topic of the following section, which
is the relationship between localization of fields and their 4D
effectiveness.

\section{Localization of Fields and Energy Cutoff}
There are two views towards fields in extra dimensions: fields can
live on a brane and do not have any freedom along the extra
dimensions; or higher dimensional fields can be localized on the
brane through some kind of mechanism. In the second view we need to
find out the mechanism about how fields are localized. The idea of
localizing fields to a topological defect originated since the early
1980s [16$\sim$19], and serves lots of different purpose such as
symmetry breakings [20, 21], fermion mass hierarchy [22, 23] and
proton decay [23]. Here we will follow the method used in [24] of
how to localize fermions on a brane.

In the simple model in [24] it is assumed that there is only one
single extra dimension parameterized by $z$. There is a 5D real
scalar field $\varphi$ whose action is

\begin{equation} \label{eps}
S_{\varphi}=\int d^{4}x\: dz\: [\frac{1}{2} (\partial
_{A}\varphi)^{2}-V(\varphi)]
\end{equation}\\
where  $A$ denotes all five coordinates. The scalar potential
$V(\varphi)$ has a double-well shape with two degenerate minima at
$\varphi=\pm v$. There exists a classical solution  $\varphi_c(z)$
which is dependent on $z$ only and can serve as a domain wall
separating two classical vacua. Introducing the Yukawa type
interaction between fermions and the scalar field $\varphi$, the
five-dimensional action for fermions can be written as

\begin{equation} \label{eps}
S_{\Psi}=\int d^{4}x\: dz\: (i\, \overline{\Psi}\Gamma^{A}\partial
_{A}\Psi-\lambdaup\overline{\Psi}\Psi)
\end{equation}\\
where $\Psi$  is the fermion field and
$\Gamma^{\mu}=\gammaup^{\mu},\: \mu=0,1,2,3;\;
\Gamma^{z}=-i\gammaup^{5}$ with $\gammaup^{\mu}$ and $\gammaup^{5}$
being the usual Dirac matrices. Then the corresponding 5D Dirac
equation is

\begin{equation} \label{eps}
i\Gamma^{A}\partial _{A}\Psi-\lambdaup\varphi_c(z)\Psi=0
\end{equation}

There exists a four-dimensional left-handed zero mode

\begin{equation} \label{eps}
\Psi_{0}=e^{\textstyle -\int_{0}^{z}dz'
\lambdaup\varphi_c(z')}\psiup_{L}(p)
\end{equation}\\
where $\psiup_{L}(p)$  is the usual solution of the 4D Weyl
equation. We can see the zero mode is localized near $z=0$, i.e., at
the domain wall. It is this zero mode which can be described as our
usual four-dimensional matter, who can acquire small masses through
some other mechanisms. This can also be interpreted as the
four-dimensional effectiveness of an effective field theory, since
interactions between zero modes can only produce zero modes again at
low energies.

There are other massive modes of the 4D fermions. The
five-dimensional fermions have a mass in the scalar field vacua
$M=\lambdaup v$. Besides the zero mode, 4D fermion field have a
continuum part of the spectrum starting at $M$, which correspond to
5D fermions which are not bound to the domain wall. Zero modes
interacting at high energies will produce these continuum modes, and
the effect of the higher dimensional part of the field will show up.
This is where the four-dimensional description of the theory loses
its effectiveness. Therefore $M$  can be regarded as an energy level
under which our theory is effectively four-dimensional, and the two
usual meanings of the term ``effectiveness'' are related here.

\section{Conclusion}
The relationship between low-energy effectiveness and
four-dimensional effectiveness can be rather complicated, and we
only provide a first sight on this question. By ignoring lots of
important problems, we try to focus on the main topic, and argue
that this relationship between the two effectiveness does exist, at
least technically. Further work need to be done in order to
investigate the specific relationship between energy cutoff and
localization of higher dimensional fields. It is also crucial to
clarify the concept of renormalizaion in this relation about
effective field theory. An important work about this question can be
found in [14].

\end{document}